\documentclass[a4paper,showpacs,twocolumn,superscriptaddress,footinbib]{revtex4}

\usepackage{graphicx}
\usepackage{hyperref}
\usepackage[sort&compress]{natbib} % natbib for hyperrefed and sorted references; numbers: numerical citation; sort: sorted citations;
\usepackage[english]{babel}
\usepackage{amsmath}
\usepackage{amssymb}
\usepackage{placeins}

\newcommand{\eref}[1]{Eq.~(\ref{#1})}
\newcommand{\fref}[1]{Fig.~\ref{#1}}

\newcommand{\TUD}{Theoriezentrum, Institut f\"ur Kernphysik,
  Technische Universit\"at Darmstadt, %Schlossgartenstrasse 2, 
  64289 Darmstadt, Germany} 

\begin{document}
\title{Gluonic three-point correlations in pure Landau gauge QCD}
\date{\today}
\author{Adrian Lorenz Blum}
\affiliation{\TUD}
% \email[Electronic address: ]{adrian@theorie.ikp.physik.tu-darmstadt.de}
\author{Markus Q. Huber}
\affiliation{\TUD}
% \email[Electronic address: ]{markus.huber@physik.tu-darmstadt.de}
\author{Mario Mitter}
\affiliation{Institut f\"ur Theoretische Physik,
 Ruprecht-Karls-Universit\"at, %Philosophenweg 16, 
 69120 Heidelberg, Germany}
% \email[Electronic address: ]{mario.mitter@thphys.uni-heidelberg.de}
\author{Lorenz von Smekal}
\affiliation{\TUD}
\affiliation{Institut f\"ur Theoretische Physik, Justus-Liebig-Universit\"at, 35392 Gie\ss en, Germany}
% \email[Electronic address: ]{lorenz.smekal@physik.tu-darmstadt.de}
\pacs{12.38.Aw, 14.70.Dj, 12.38.Lg}%gluons; General properties of QCD (dynamics, confinement, etc.); Other nonperturbative calculations

\begin{abstract}
We report on the first self-consistent solution of the Dyson-Schwinger
equation (DSE) for the three-gluon vertex. Based on earlier results for the
propagators which match data from lattice Monte-Carlo simulations, we
obtain results for the three-gluon vertex that are in very good
agreement with available lattice data likewise. Feeding these results
back into the propagator DSEs leads to some changes especially in the
gluon propagator.  These changes allow us to assess previously used
models for the three-gluon vertex and to systematically estimate the
influence of neglected two-loop diagrams with four-gluon
interactions. In the final step, a full iterative solution to the
coupled DSEs of pure Landau gauge QCD without quarks is then obtained
for the first time in an extended truncation which now dynamically
includes the complete set of three-point vertex functions. 
\end{abstract}

\maketitle

{\bf Introduction.}
The correlation functions of quantum chromodynamics (QCD) are the fundamental building blocks for hadron phenomenology
\cite{Eichmann:2013afa} and strong-interaction matter studies
 based on functional continuum methods \cite{Pawlowski:2010ht}. Built
 on a solid understanding of the pure gauge
 theory's vacuum correlations
 \cite{Alkofer:2000wg,Lerche:2002ep,Pawlowski:2003hq,Fischer:2008uz,Huber:2012kd},
 there has 
 recently been considerable progress in extensions to finite temperature
\cite{Fischer:2010fx,Maas:2011ez,Fister:2011uw}, to  
including dynamical quarks
\cite{Braun:2009gm,Fischer:2011mz,Aouane:2012bk}, and to finite baryon density with all three
light quark flavors included \cite{Fischer:2012vc}.  Via
corresponding calculations of Polyakov-loop potentials
\cite{Braun:2007bx}, which have recently also included 
unquenching \cite{Fister:2013bh,Haas:2013qwp} and quark matter effects
\cite{Fischer:2013eca}, they also provide input for
Polyakov-loop extended quark \cite{Hell:2011ic} and quark-meson models
\cite{Herbst:2013ufa}.

In this paper we go back to the foundations and consider pure Landau
gauge QCD without quarks, for which the correlation functions have been
intensely studied within a variety of approaches. These range from
 Monte-Carlo simulations on discrete space-time lattices
to functional continuum methods such as Dyson-Schwinger equations
(DSEs) or Functional Renormalization Group studies. Thereby, good
qualitative agreement has been achieved since lattice sizes have
become large enough to access the deep infrared (IR), far below the
scale of QCD, $\Lambda_\mathrm{QCD}$, in simulations
\cite{Cucchieri:2007md,Cucchieri:2008fc,Sternbeck:2007ug,Bogolubsky:2009dc}.
At intermediate momenta, of the order of $\Lambda_\mathrm{QCD}$, where much of
the non-perturbative dynamics relevant to hadron physics happens,
however, there are still some quantitative discrepancies. 
Especially in view of applying functional continuum methods to
strong-interaction matter at finite baryon density, where the fermion
sign problem is impeding direct lattice simulations, it is worthwhile 
to resolve these discrepancies. The QCD vacuum correlations thus serve as an
important benchmark before the distinctive feature of functional methods to
be readily extensible to finite baryon density can fully and reliably
be exploited. 

Moreover, a key role in hadron physics and finite density applications
is increasingly being played by 3-- and higher $n$--point vertex
functions. Even for the 3--point vertex functions, however, lattice
data is rather limited, see
\cite{Kizilersu:2006et,Cucchieri:2008qm,Ilgenfritz:2006he}.   
This is to some extent due to their more complicated kinematics. 
Typically, lattice data
has therefore so far only become available for very restricted kinematical
configurations. While such restricted data provides valuable constraints,
functional methods can also fill this gap and yield kinematically
complete descriptions.  

On the other hand, the infinite sets of functional equations for correlation
functions require truncations. In the past this basically always meant
that model input was used for the 3--point vertex functions to
self-consistently solve non-linear functional equations for the
propagators
\cite{vonSmekal:1997is,vonSmekal:1997vx,Aguilar:2008xm,Fischer:2008uz,Pennington:2011xs,LlanesEstrada:2012my}. 
While such 2--point complete truncations are nowadays being extended
into the complex invariant momentum plane, e.g., for direct calculations 
of the corresponding spectral functions \cite{Strauss:2012dg}, the
fully self-consistent inclusion of dynamic 3--point vertex functions has
only started very recently \cite{Huber:2012kd,Hopfer:2013via}.
Despite constituting a major conceptual breakthrough the structurally
simplest of all QCD vertex functions, the ghost-gluon vertex,
was thereby shown to only have a minor quantitative influence on the
propagators \cite{Huber:2012kd} as predicted 
\cite{Lerche:2002ep} and confirmed in \cite{Aguilar:2013xqa}.
In contrast, it is usually argued that the three-gluon vertex plays a
crucial role in the mid-momentum regime around $\Lambda_\mathrm{QCD}$. 
The limited lattice data available for this vertex
\cite{Cucchieri:2008qm}, however, left considerable room for
speculations and models that had to be used in the past.
Recently also perturbative calculations with a Curci-Ferrari mass term
were done \cite{Pelaez:2013cpa} that describe the qualitative features
of the vertex quite well.

In this paper we present the next major step which is to also include the
three-gluon vertex fully self-consistently and dynamically in a DSE
solution for the pure gauge theory which is thus now 3--point
complete for the first time. 
Before that, however, we first describe a standalone solution to a
truncated three-gluon vertex DSE based on input propagators from
\cite{Huber:2012kd} that are in very good agreement with lattice
data. The fact that this standalone solution is 
then in turn consistent with the available lattice data for the three-gluon
vertex confirms the validity of our truncation of this DSE.
Feeding the lattice-consistent three-gluon vertex back into the
propagator DSEs serves to demonstrate to what extent the previously
used model vertex \cite{Huber:2012kd} effectively includes
contributions from the neglected two-loop diagrams in the gluon
propagator DSE. The final step then is the fully iterated solution to
the 3--point complete set of propagator {\em and} vertex DSEs, now
based on a four-gluon vertex model.
While the iteration has some effect on the three-gluon vertex, it
hardly changes the propagators anymore which
is encouraging evidence of convergence of this
type of vertex expansion.

\medskip

{\bf Calculational scheme.}
The general setup follows that of Ref.~\cite{Huber:2012kd} where the
coupled system of ghost-gluon vertex and propagator DSEs was solved. The
Landau gauge gluon and ghost propagators are parameterized by two
invariant functions (color indices suppressed),
\begin{align}
D^{A}_{\mu\nu}(p)=P_{\mu\nu}(p)\frac{Z(p^2)}{p^2} & \;\;\mbox{and} 
\quad D^{c}(p)=-\frac{G(p^2)}{p^2},
\end{align}
where $P_{\mu\nu}(p)$ is the transverse projector. In the Landau gauge, 
the relevant transverse part of the three-gluon vertex can be written
in terms of four independent Lorentz tensors. Including a complete basis
for this tensor structure in the three-gluon vertex DSE, one can show, however, 
that the transverse part of the tree-level structure provides the
dominant contribution to the full three-gluon vertex
\cite{Eichmann:2014xya}. In our present study, we therefore maintain only
this tree-level structure in the ansatz, 
\begin{align}
 &\Gamma^{A^3,abc}_{\mu\nu\rho}(p,q,k)=\nonumber\\
&i\,g\,f^{abc}D^{A^3}(p^2,q^2,\alpha)\left((q-p)_\rho g_{\mu\nu} + perm. \right),
\label{eq:3glv-ans}
\end{align}
where $\alpha$ is the angle between momenta $p$ and $q$. To
project the three-gluon vertex DSE onto this structure we contract it
with three transverse projectors and a tree-level three-gluon vertex,
for which $D^{A^3}\equiv 1$, as it is also done in lattice
calculations \cite{Cucchieri:2008qm}. One advantage of this procedure
is that the same projection occurs in the gluon loop of the gluon
propagator DSE. Consequently, the error induced in the gluon DSE by
this restriction can be quantified from comparing the so projected
vertex DSE results with analogously projected lattice data.  

\begin{figure}[tb]
  \includegraphics[width=\columnwidth]{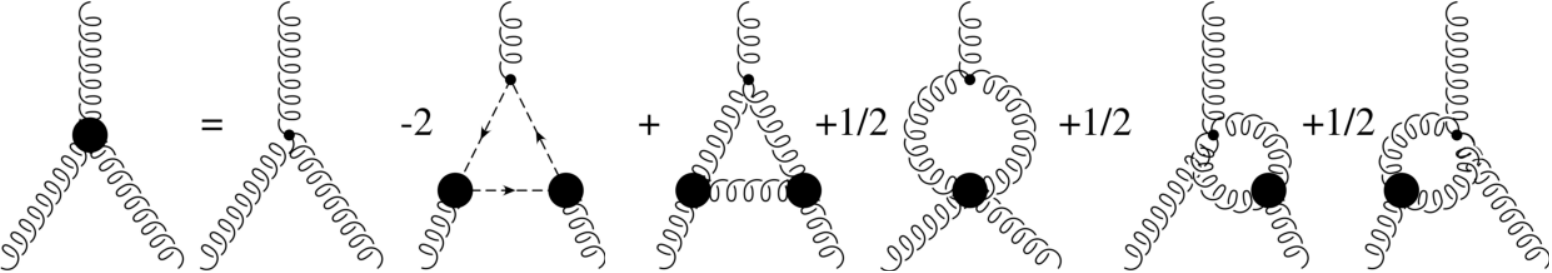}
  \caption{\label{fig:3g-DSE}Truncated three-gluon vertex
    DSE with (dashed) ghost and (wiggly) gluon triangles, plus
    so-called swordfish diagrams with 4--point interactions. Solid black disks represent dressed
    vertices. Propagators inside loops are also dressed.} 
\end{figure}

The full three-gluon vertex DSE, whose diagrammatic form can be found
in Ref.~\cite{Huber:2012zj} for example, is truncated by discarding
all explicit two-loop diagrams together with a diagram that contains
an irreducible ghost-gluon scattering kernel without tree-level
counterpart. The resulting truncation is shown diagrammatically  
in \fref{fig:3g-DSE}. It is complete at leading order in the
ultraviolet (UV). Moreover, it also includes the IR dominant
contribution given by the ghost triangle, so that truncation errors
should manifest themselves only in the mid-momentum region.
To obtain a Bose symmetric result, the DSE is finally symmetrized
by averaging over all three possible positions of the bare vertex in the diagrams.

For renormalization we use the \textit{MiniMOM} scheme \cite{vonSmekal:2009ae},
i.e., minimal subtraction of the ghost-gluon vertex (which entails
$\tilde{Z}_1=1$ in Landau gauge) combined with momentum subtraction
for the propagators. The renormalization constant of the three-gluon vertex is 
then fixed by its Slavnov-Taylor identity, 
$Z_1=Z_3/\tilde{Z}_3$, where $ Z_3$ and $\tilde Z_3$ are the
renormalization constants of gluon and ghost fields,
respectively.
$Z_1$ and $Z_4$ factors also come with the tree-level vertices in
gluon loops. To reproduce correct anomalous dimensions,
we have to replace them by momentum dependent factors there
\cite{vonSmekal:1997vx}. Our construction of $Z_1$ for this
renormalization group improvement is described in
Ref.~\cite{Huber:2012kd}. For $Z_4$ we use analogously,
\begin{align}\label{eq:4g-RG}
 Z_4 \rightarrow D^{A^4}_{RG}(p, q, r, s) =
 G\left(\bar{p}^2\right)^{\alpha_{4g}}
 Z\left(\bar{p}^2\right)^{\beta_{4g}}, 
\end{align}
where $\bar{p}^2=(p^2+q^2+r^2+s^2)/2$. The exponents $\alpha_{4g}$ and
$\beta_{4g}$ are then determined from
the leading anomalous dimension of the four-gluon vertex,
$\gamma_{4g}=2/11$, and
from the requirement that the vertex approaches a constant value in
the IR. These two conditions together yield $\alpha_{4g}=-8/9$ and
$\beta_{4g}=0$. 

Note that there is no freedom in the subtraction of the three-gluon vertex
DSE because this would in general violate the Slavnov-Taylor identity
and hence be inconsistent with the {MiniMOM} scheme. Its overall
strength can therefore not be adjusted manually by
renormalization. In general one observes, however, that the iteration
of the three-gluon vertex DSE with fixed propagator
input, roughly comparable to lattice data, does not converge once
three-gluon interactions of a certain strength build up \footnote{We
  thank G.~Eichmann and R.~Williams for bringing this to our
  attention, for numerical cross\--checks, and for clarifying
  discussions on how to resolve the issue.}. Because of cancellations between the 
gluon triangle, where these enter quadratically, and the swordfish
diagrams, with four-gluon interactions, this can at present only be
avoided by using a sufficiently strong four-gluon vertex as model input
\cite{Eichmann:2014xya}. Especially its strength in the mid-momentum regime 
is thereby important for the balance between gluon triangle and swordfish
diagrams. This suggests that the neglected UV-subleading contributions and
tensor structures might have a similar effect in the full DSE. It is
thus in line with our general strategy for the vertex
expansion that such higher-order effects are compensated by the model
input for the 4--point interactions used to close the 3--point complete
system of DSEs. 
The situation is analogous to that in previous 2--point
complete truncations, where models for the three-gluon vertex were
also partially constrained by the convergence of the gluon propagator DSE
solution. 

\begin{figure*}[tb]
  \includegraphics[width=0.32\textwidth]{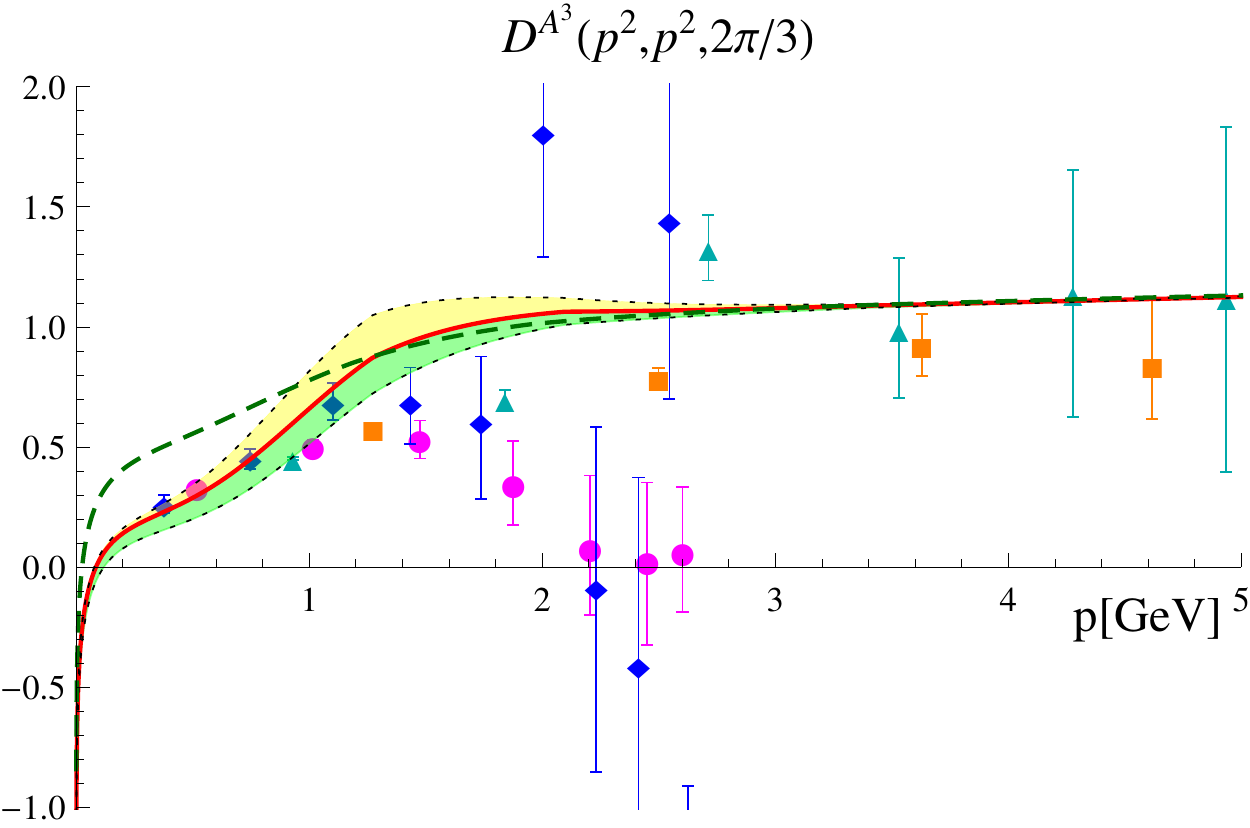}
  \includegraphics[width=0.32\textwidth]{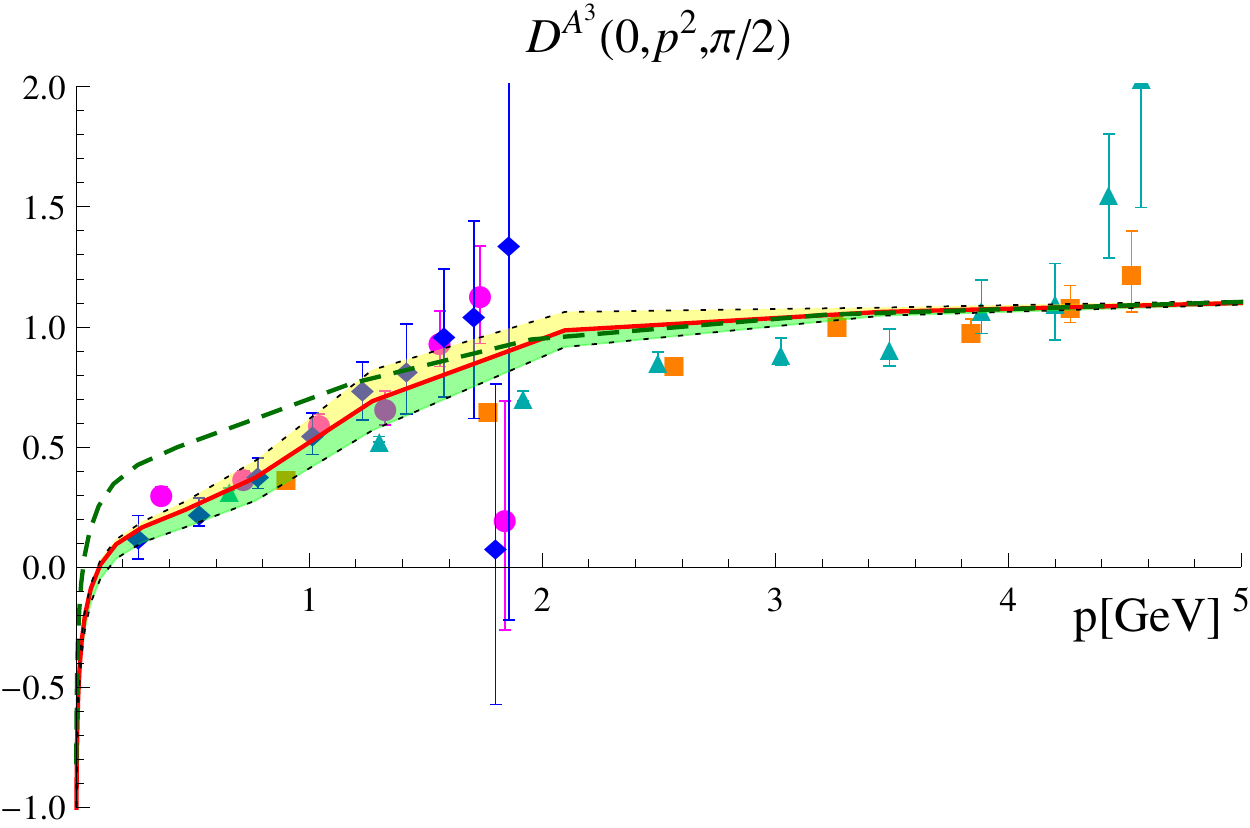}
  \includegraphics[width=0.32\textwidth]{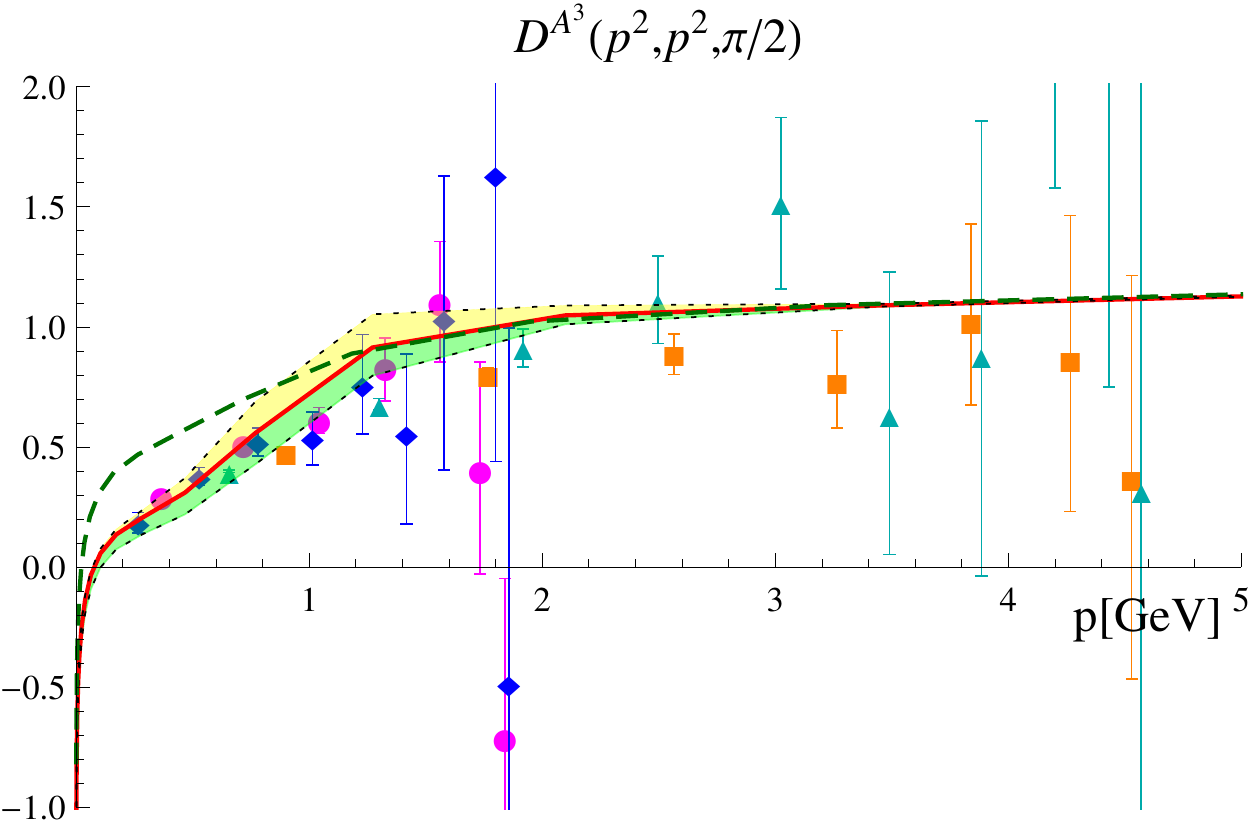}
  \caption{\label{fig:3g_singleScale}Three-gluon vertex dressing
    function with restricted kinematics (see legends) for comparison
    with lattice data where different colors/symbols refer to different values of $\beta\in \{2.2, 2.5\}$ and different lattice sizes $1.4$ fm $<L<4.7$ fm, see \cite{Cucchieri:2008qm} for details. Solid red line:
    standalone solution with $a=1.5$ and $b= 1.95$ GeV$^2$. Upper
    (yellow) band: variation with $b$ down to $ 1.46$ GeV$^2$. Lower
    (green) band: strengths up to $a=2$.
    Green dashed line: solution to fully coupled system ($a=1.5$,
    $b=1.94$ GeV$^2$).}  
\end{figure*}

\begin{figure}[b]
  \includegraphics[width=\columnwidth]{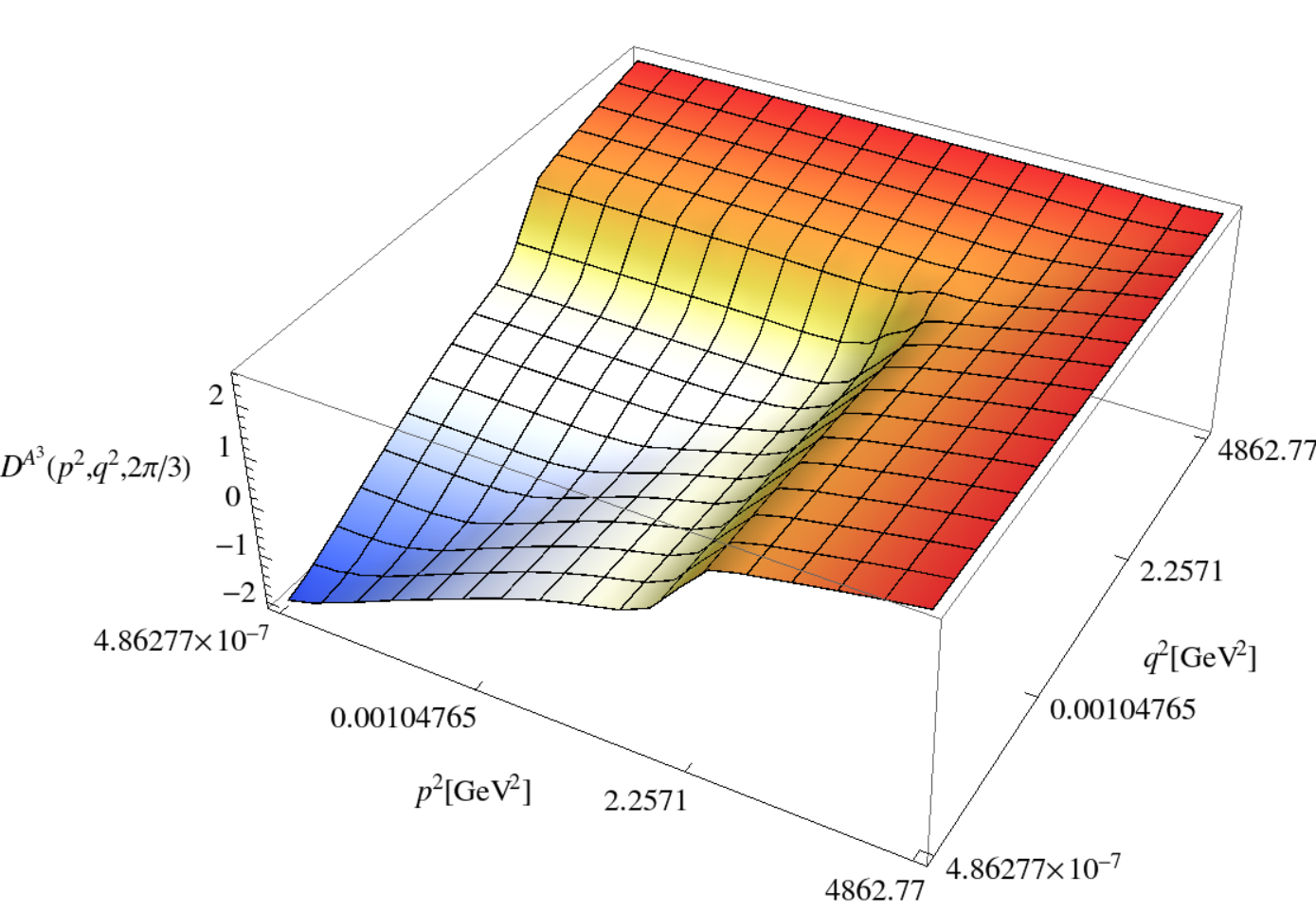}
  \caption{\label{fig:3g_3d}Three-gluon vertex dressing function for
    $\alpha = 2\pi/3$.
    }

\vspace*{-.2cm}
\end{figure}

We again use a tree-level ansatz for the Lorentz and color structure
of the dressed four-gluon vertex. To enhance its low and mid-momentum
strength as compared to the form in \eref{eq:4g-RG} we use a two
parameter ansatz for its dressing,
\begin{align}\label{eq:4g-model}
D^{A^4}(p, q, r, s) =(a \tanh(b/\bar{p}^2)+1)\, D^{A^4}_{RG}(p, q, r, s),
\end{align}
where $a$ determines the additional IR interaction strength and $b$
the momentum scale of its onset. Qualitatively, such an enhancement is
in fact in agreement with a first exploratory study of the four-gluon
vertex function \cite{Kellermann:2008iw}.

As in Ref.~\cite{Huber:2012kd}, where further technical details are
found, the program \textit{DoFun} \cite{Huber:2011qr,Alkofer:2008nt}
was used to derive the DSEs and the \textit{CrasyDSE} framework
\cite{Huber:2011xc} for their solution.

\medskip

\begin{figure}[tb]
  \includegraphics[width=\columnwidth]{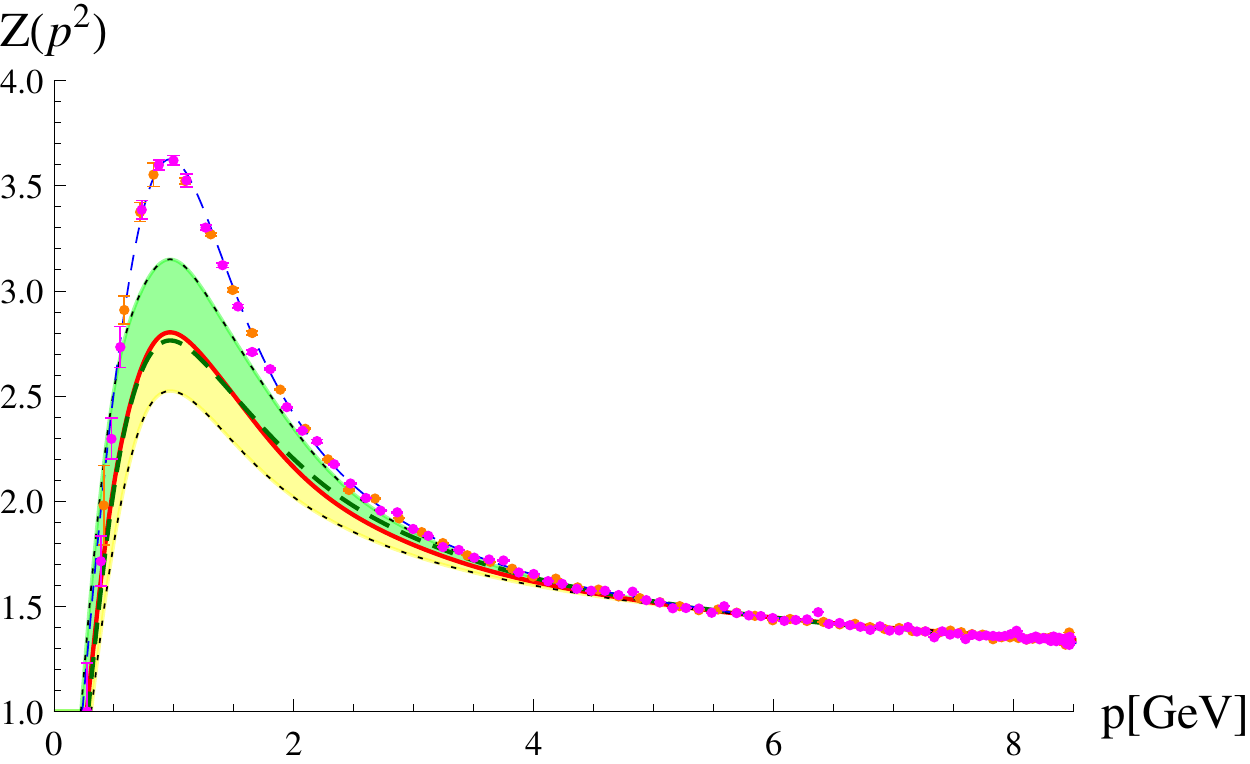}
  \caption{\label{fig:gl}Gluon dressing function from
    \cite{Huber:2012kd} (dashed blue), with lattice data from
    \cite{Sternbeck:2006rd}, compared to analogous calculations with
    the three-gluon vertices shown as the bands around the solid red
    line in \fref{fig:3g_singleScale} and the same color coding
    here. Iterating the full 3--point complete set of DSEs then
    changes the center of the band (solid red) into the dashed green line.}
\end{figure}

{\bf Standalone three-gluon vertex.}
With the ghost-gluon vertex and the propagators from \cite{Huber:2012kd}
as fixed input which agrees with lattice data very well, see, e.g.,
the dashed blue line for the gluon propagator in \fref{fig:gl}, the
calculation of the three-gluon vertex serves as a test of its
truncated DSE with simplified tensor structure and the four-gluon
vertex model (\ref{eq:4g-model}). In  \fref{fig:3g_singleScale} we
compare the resulting dressing function $D^{A^3}(p^2,q^2,\alpha)$
defined in \eref{eq:3glv-ans} for the symmetric momentum configuration
$k^2=p^2=q^2$ (left)  and for two orthogonal configurations  with $p\cdot q=0$,
for $k^2 = p^2$ (middle) and $q^2=p^2$ (right), to the lattice data of
Ref.~\cite{Cucchieri:2008qm}. A rather good description of the lattice
data is obtained for an IR strength parameter $a\approx 1.5$ with
an onset around $b\approx 2$ GeV$^2$ in the four-gluon vertex
model. Varying its strength and onset by about 30\% leads to the bands
used in the figure to indicate the sensitivity to these model
parameters. The four-gluon vertex model thus appears to compensate the
mid-momentum contributions from neglected diagrams in the tree-level
projected three-gluon vertex DSE quite well.

As already observed in \cite{Huber:2012kd}, our results for the
three-gluon vertex function change sign at very low momenta. 
The position of this zero crossing in our calculations for the momentum
configurations of \fref{fig:3g_singleScale} varies roughly between 80
and 100 MeV, typically with a 20\% variation over the bands. 
Our larger value, which is obtained for the configuration in the middle,
is thereby reasonably close to a previous estimate of about $130$ MeV
\cite{Aguilar:2013vaa}. With the available lattice sizes this 
zero crossing has not yet been accessible by Monte-Carlo
simulations in four dimensions. It has been observed, however, by
simulations on larger lattices in two \cite{Maas:2007uv} and three
dimensions \cite{Cucchieri:2008qm}, where it was confirmed by DSE studies
\cite{Campagnari:2010wc, Huber:2012zj}.  

For even lower momenta our results for the three-gluon vertex function
furthermore show a logarithmic behavior with signs of a
divergence at vanishing momenta. For the momentum configuration in
the middle of \fref{fig:3g_singleScale}, such a logarithmic divergence 
has also been predicted in \cite{Aguilar:2013vaa}. 
It agrees with the general arguments of Ref.~\cite{Alkofer:2008jy},
and it has been seen explicitly in a recent perturbative calculation with
Curci-Ferrari mass term as well \cite{Pelaez:2013cpa}. 

Strictly speaking, the lattice data shown here was obtained for the
pure $SU(2)$ gauge theory, while our DSE calculations refer to $SU(3)$. At
the present accuracy level, this difference is not significant,
however. The renormalized propagators of the two basically
coincide \cite{Sternbeck:2007ug}. For completeness we provide
corresponding DSE results for $SU(2)$, which indeed compare equally
well with the lattice data, in the Supplemental Material.

A typical example of how our results extend the lattice data in
\fref{fig:3g_singleScale}, here with $\alpha = 2\pi/3$ (left),
to general kinematics is shown in \fref{fig:3g_3d}. Analogous
$SU(2)$ results and further examples are given in the Supplemental
Material.

\medskip

{\bf Gluon propagator.}
As mentioned above, the way we project the three-gluon vertex DSE onto
the tree-level structure (\ref{eq:3glv-ans}) leads to the same
structure that also occurs inside the gluon loop of the gluon
propagator DSE. Using in this DSE a three-gluon vertex for which this
same structure resembles lattice data therefore practically eliminates
the effects of other tensor structures on the gluon propagator, which is
shown in \fref{fig:gl}.
Solving its DSE with the three-gluon vertices shown as the bands in
\fref{fig:3g_singleScale} reduces the result of \cite{Huber:2012kd}
(dashed blue), i.e., the input for these  vertex DSE
solutions, to the corresponding bands around the solid red line here.  
Since these three-gluon vertices agree within errors with the  
lattice data, the missing strength of the gluon propagator in the
mid-momentum regime has to come from the neglected two-loop
diagrams which hence deserve further study. For first results see  
\cite{Bloch:2003yu,Mader:2013ru}. Ghost propagator and $SU(2)$ results
are given in the Supplemental Material.

\medskip 

{\bf Full 3--point complete solution.}
We have seen that with proper input our standalone three-gluon vertex
DSE solution agrees well with lattice data. Using this solution in the
gluon propagator DSE exposes missing contributions there. The
resulting gluon propagator decreases at mid-momentum and no longer
agrees with lattice data. If we feed this result back into the vertex
DSE, it is thus to be expected that its agreement with lattice data
deteriorates, likewise. This is indeed the case as seen in
\fref{fig:3g_singleScale} where the dashed green lines show the 
iterated and converged solution for the three-gluon vertex from the
3--point complete system of propagator and 3--point vertex DSEs. Apart
from the expected deviations, these fully self-consistent results
are otherwise stable, however. 

In particular, it is quite reassuring for the convergence of this kind
of vertex expansion that the propagators remain almost unaffected by
these deviations in the three-gluon vertex as can be seen, for
example, in the gluon propagator upon comparing the solid red line of  
\fref{fig:gl}, from the lattice consistent vertex, with the dashed
green fully iterated result, corresponding to the fully
iterated dashed green vertex result in \fref{fig:3g_singleScale}.

The ghost propagator and ghost-gluon vertex are 
both affected very little by the inclusion of the three-gluon
vertex in the 3--point complete iteration and are not shown here.

\medskip

\begin{figure}[tb]
  \includegraphics[width=\columnwidth]{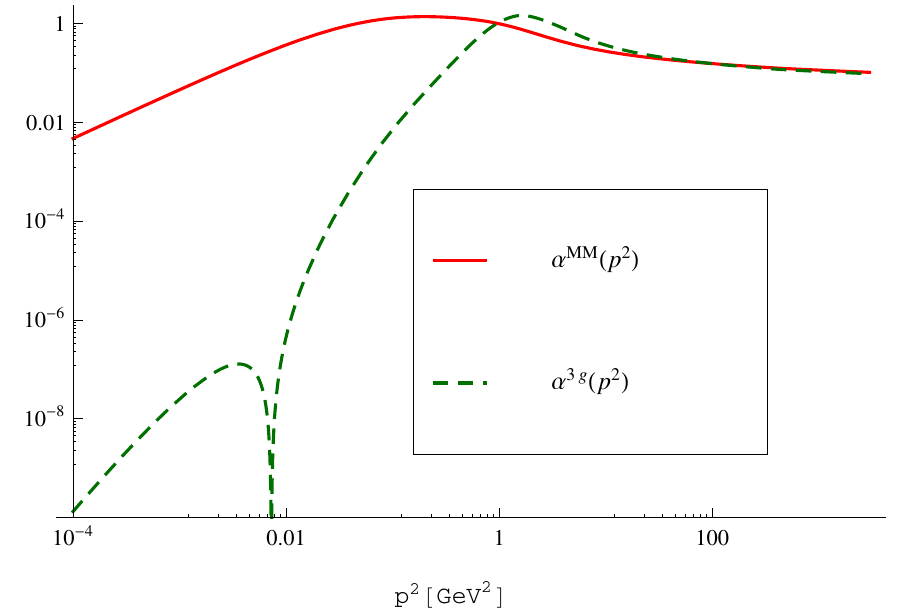}
  \caption{\label{fig:couplings}Comparison of couplings as defined via
    ghost-gluon (solid red) and three-gluon (dashed green) vertex.}
\end{figure}

{\bf Running couplings.}
The {MiniMOM} coupling is defined by minimal subtraction of 
the ghost-gluon vertex, i.e., in Landau gauge as $\alpha^\mathrm{MM}(p^2)=\alpha(\mu^2)Z(p^2) G(p^2)^2$ \cite{vonSmekal:1997is,vonSmekal:2009ae}.
Alternatively, one could of course also define a running coupling from  
the three-gluon vertex in a symmetric MOM scheme, for example, see
Ref.~\cite{Alkofer:2004it},  
\begin{align}
  \alpha^{3g}(p^2)=\alpha(\mu^2)\frac{Z(p^2)^3 D^{A^3}\left(p^2,p^2, 2\pi/3 \right)^2}{Z(\mu^2)^3 D^{A^3}\left(\mu^2,\mu^2, 2\pi/3 \right)^2}.
\end{align}
The denominator herein, which would be unity with subtracting at
$p^2=\mu^2$ in such a symmetric MOM scheme, is used to
convert our $Z$ and $D^{A^3}$ from the MiniMOM scheme to this scheme.
For $\mu$ far enough in the perturbative regime the two couplings
must agree in the UV. 
A comparison is given in \fref{fig:couplings}. Because of the zero
crossing in the three-gluon vertex, $\alpha^{3g}(p^2)$ has
a zero at non-vanishing momentum likewise. This is not prohibited, in
general, for a renormalization group invariant dimensionless function
of a single scale which reduces to the perturbative running coupling
in the UV, but it does certainly go against the common notion of a running
coupling.  

\medskip 

{\bf Summary and conclusions.}
We have shown how a truncated DSE for the three-gluon vertex with
appropriate input and modeling of four--gluon interactions can
produce reliable results which stand the test against current lattice
data. Using these results in the gluon propagator DSE, we could clearly
identify missing contributions in the mid-momentum regime around 1 GeV
as being due to neglected two-loop diagrams therein. Our solid results
for the three-gluon vertex will help to include these diagrams in the
future. Meanwhile we have solved for the first time a coupled system
of DSEs for propagators {\em and} vertex functions  that is complete
on the level of 3--point correlations with model 4--point
interactions. The fully self-consistent results from this 3--point
complete truncation show clear signs of convergence of the underlying
vertex expansion for QCD.

\medskip

{\bf Acknowledgments.}
We would like to thank Reinhard Alkofer, Gernot Eichmann, Christian
S.~Fischer, Leonard~Fister, Axel Maas, Jan M.~Pawlowski, Andre
Sternbeck, and Richard Williams for helpful discussions. 
This work was supported by the Helmholtz International Center for FAIR
within the LOEWE program of the State of Hesse, the European
Commission, FP7-PEOPLE-2009-RG No. 249203, the Alexander von Humboldt
foundation and the BMBF grant OSPL2VHCTG.

\bibliographystyle{utphys_mod}
\bibliography{literature_3point_complete}

\FloatBarrier

% \newpage

\section*{Supplemental Material}

\begin{figure}[h]
  \includegraphics[width=0.49\textwidth]{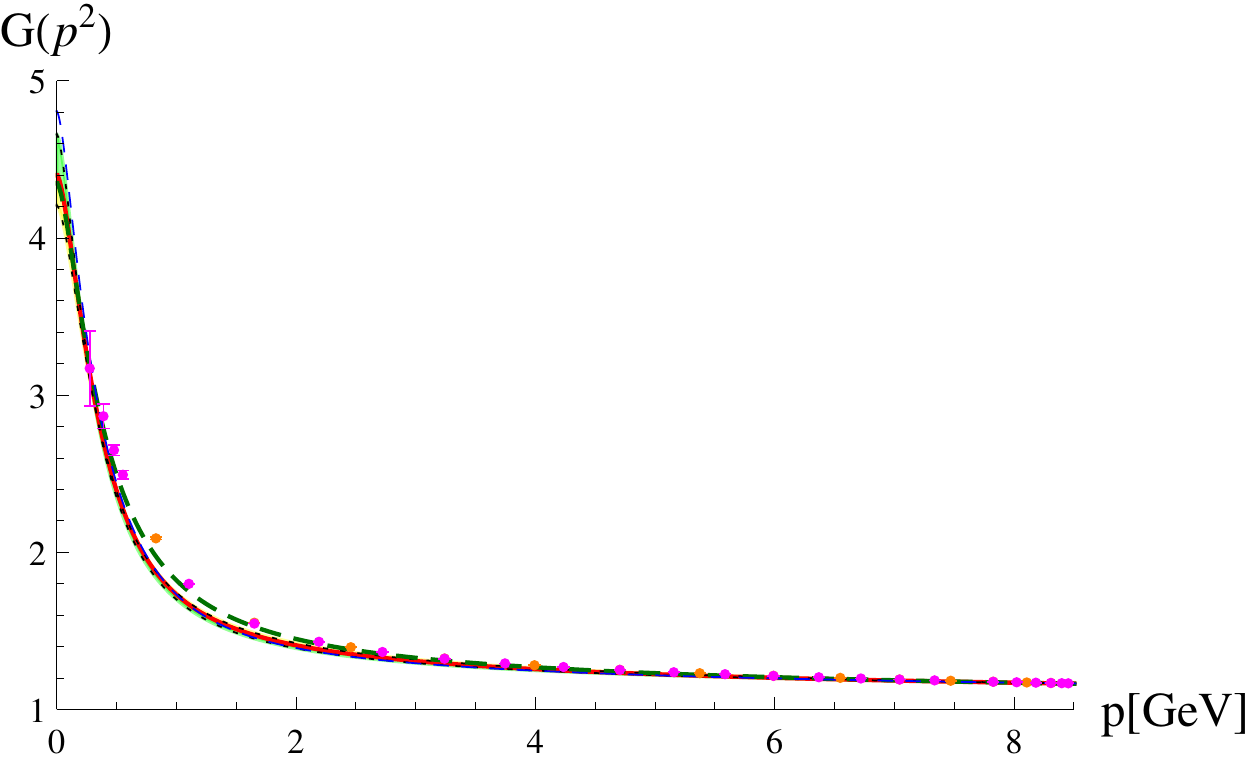}
  \caption{\label{fig:gh}$SU(3)$ ghost dressing functions: Input
    from \cite{Huber:2012kd} depicted by a dashed blue
    line. Results from standalone solution to three-gluon vertex
    DSE shown as solid red line (with hardly visible small bands
    corresponding to those in Figs.~\ref{fig:3g_singleScale} and
    \ref{fig:gl}), and iterated solution from the 3--point 
    complete truncation as dashed green line, with lattice data from
    Ref.~\cite{Sternbeck:2006rd}.} 
 \end{figure}

For completeness we show the ghost dressing function in \fref{fig:gh}. As expected, the effect of the three-gluon vertex is only minor since it enters only indirectly via the gluon propagator.

\begin{figure}[h]
  \includegraphics[width=0.49\textwidth]{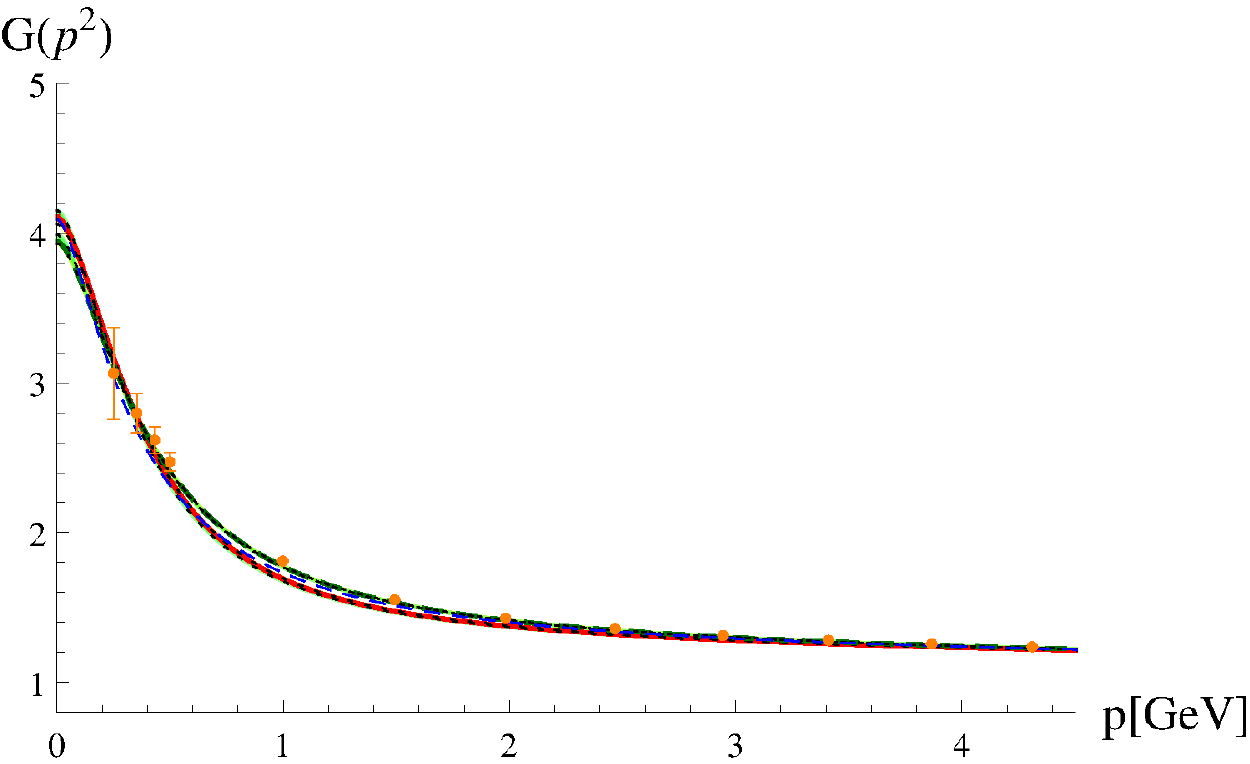}
  \caption{\label{fig:prop_gh-SU2}$SU(2)$ ghost dressing functions
    with lattice data from \cite{Sternbeck:2007ug} corresponding to those
    shown for $SU(3)$ in \fref{fig:gh}.}
\end{figure}

\begin{figure}[h]
  \includegraphics[width=0.49\textwidth]{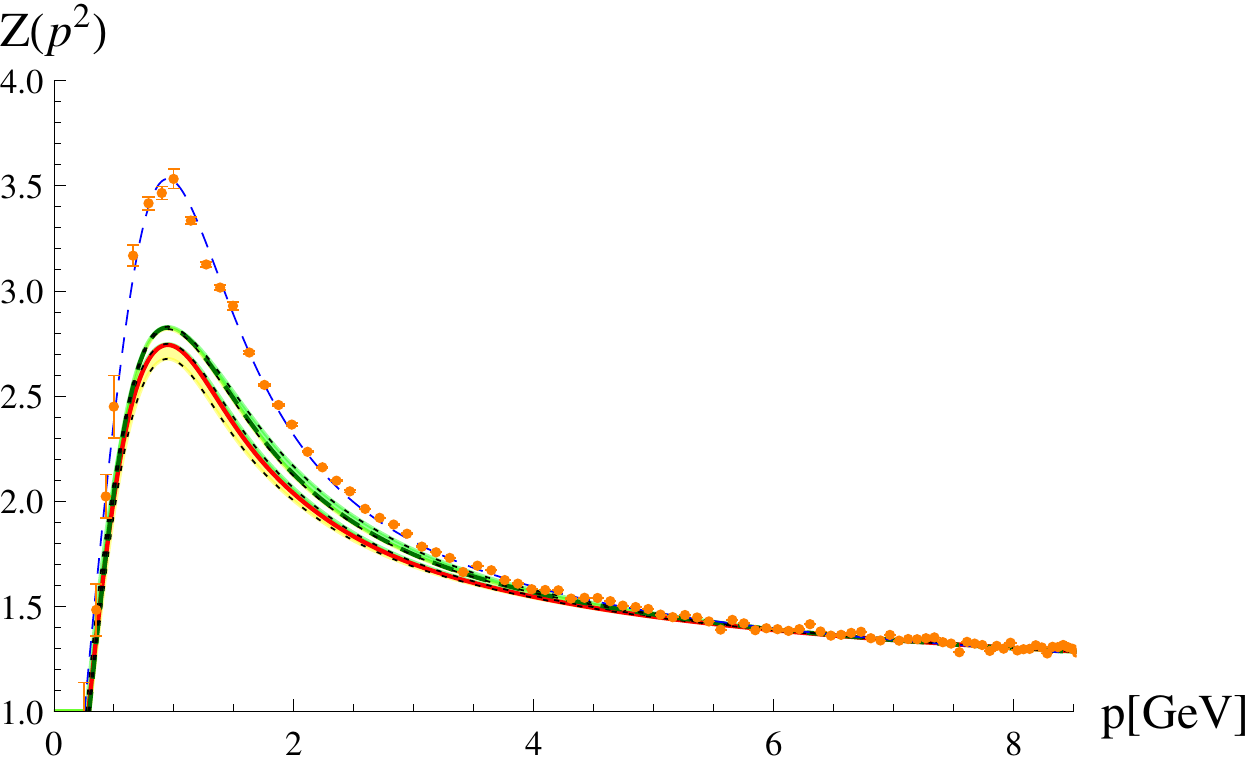}
  \caption{\label{fig:prop_gl-SU2}Gluon dressing functions as
    in \fref{fig:gl}, here for $SU(2)$: Input from \cite{Huber:2012kd}
    (dashed blue), lattice data from \cite{Sternbeck:2007ug}, results
    with standalone three-gluon vertex DSE solution  (solid red with
    bands), and  3--point complete self-consistent solution (dashed
    green).} 
\end{figure}

\begin{figure}[b]
  \includegraphics[width=0.49\textwidth]{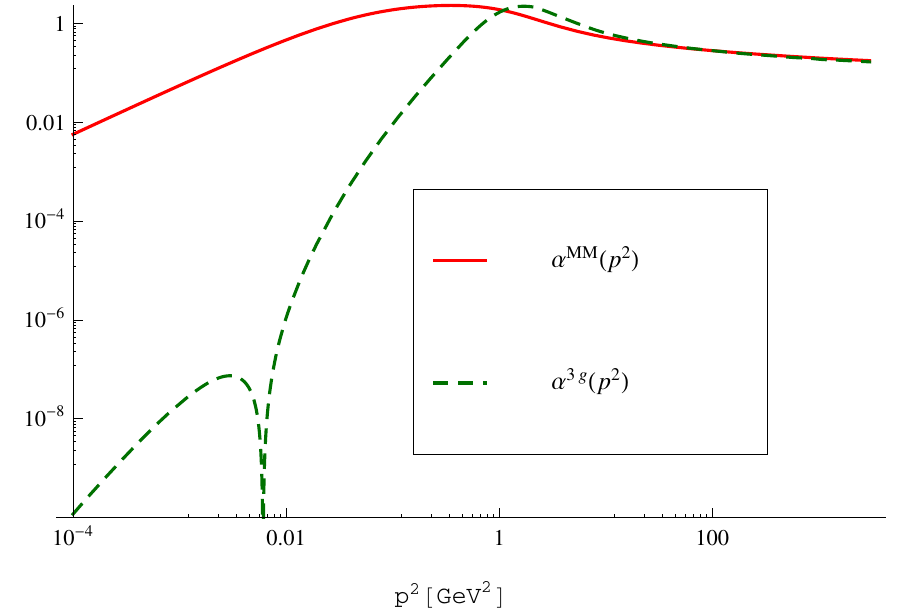}
  \caption{\label{fig:couplings-SU2}$SU(2)$ couplings obtained from
    ghost-gluon (solid red) and three-gluon (dashed green) vertices.}
\end{figure}

\begin{figure*}[tb]
  \includegraphics[width=0.32\textwidth]{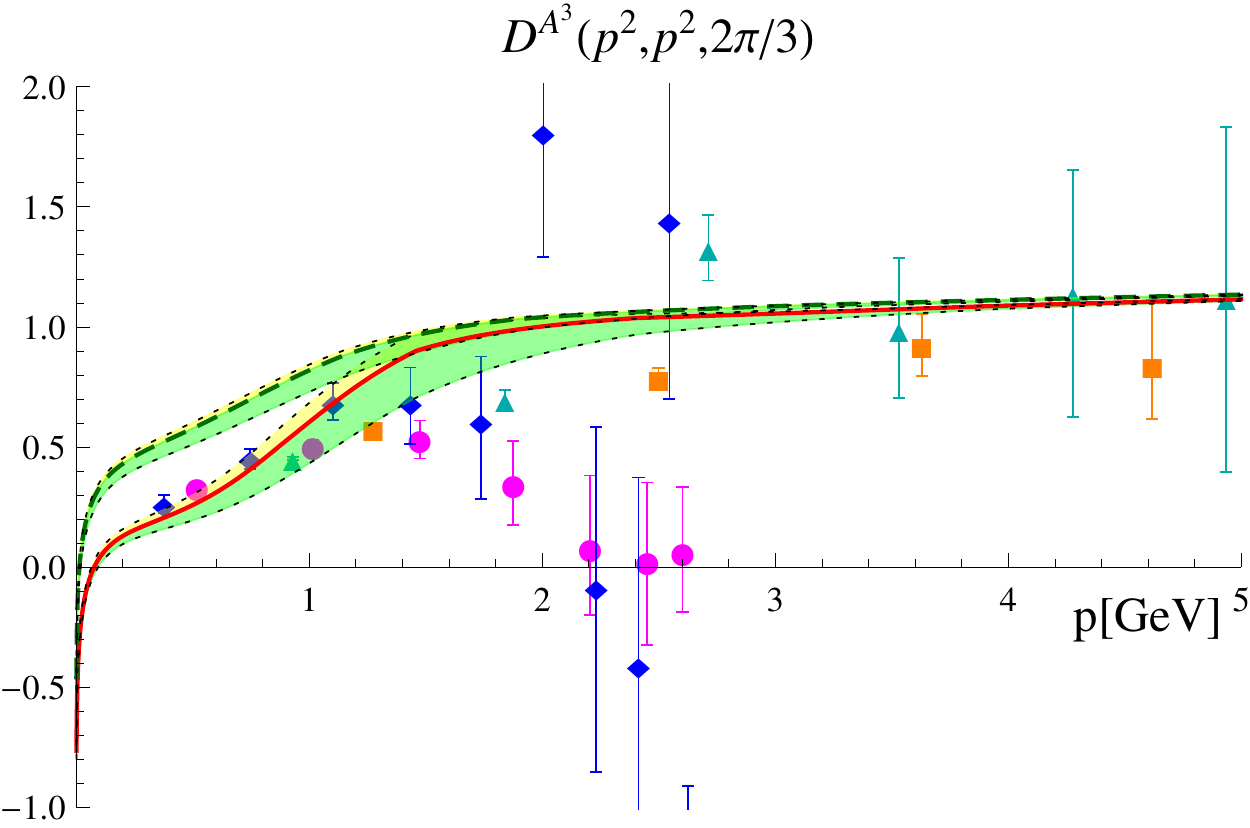}
  \includegraphics[width=0.32\textwidth]{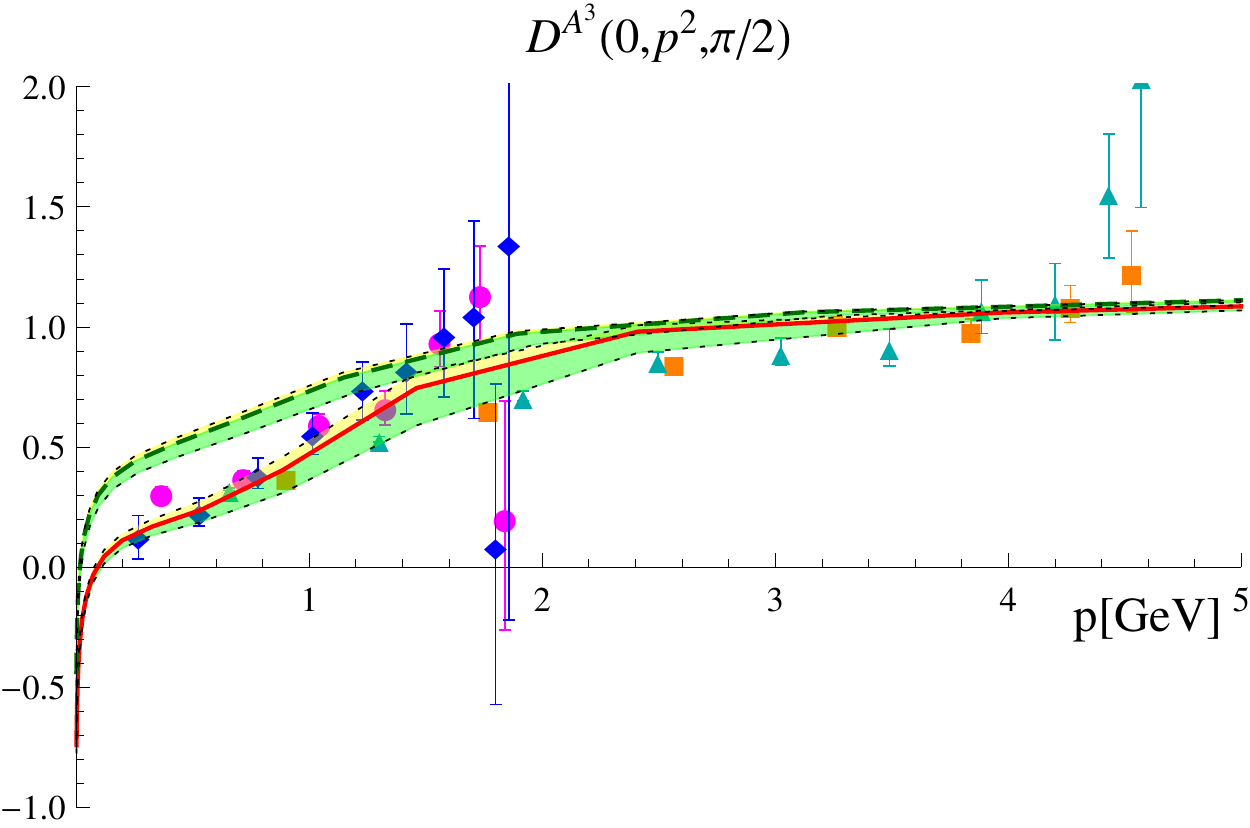}
  \includegraphics[width=0.32\textwidth]{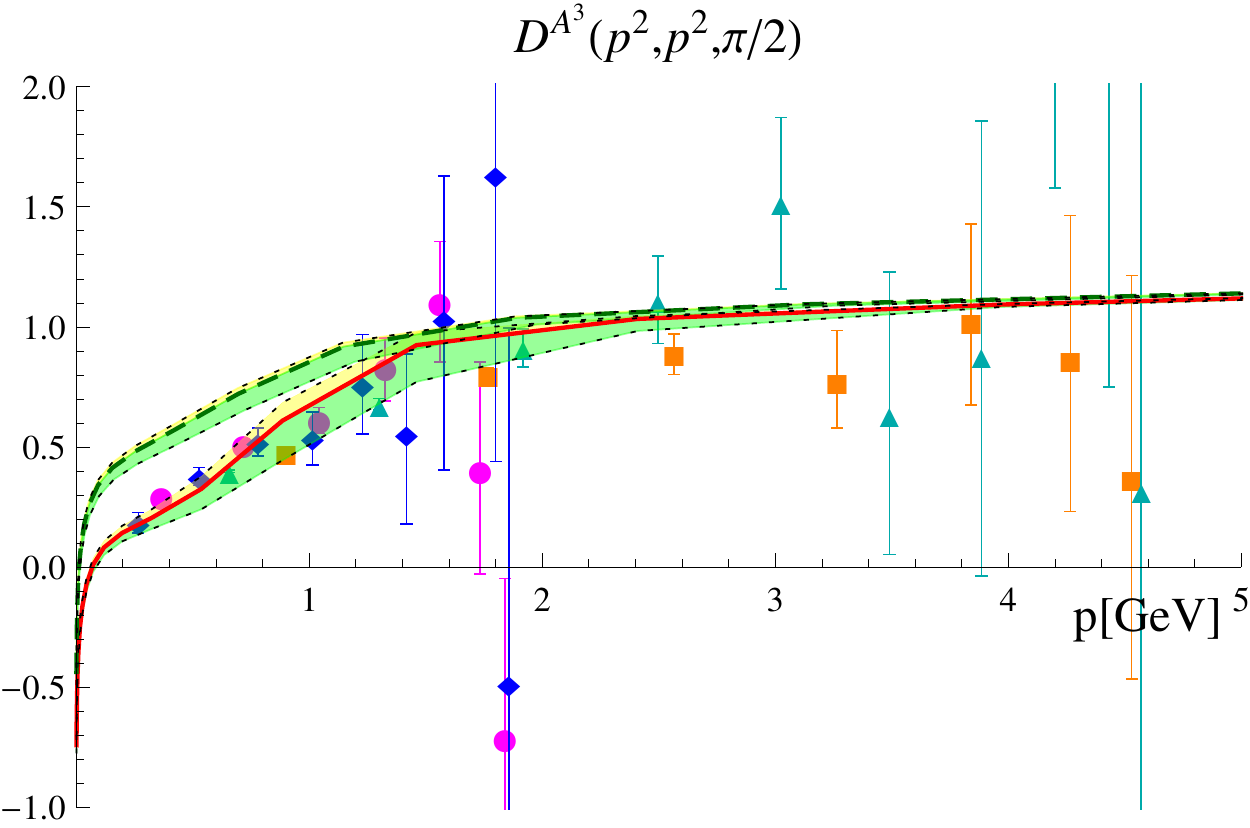}
  \caption{\label{fig:3g_singleScale-SU2}$SU(2)$ three-gluon vertex
    for three kinematical configurations (as specified above the
    plots) in comparison with $SU(2)$ lattice data from
    \cite{Cucchieri:2008qm}. Color coding as in
    \fref{fig:3g_singleScale}, solid red line: standalone vertex DSE 
    solution with  $a=1.5$ and $b=1.67$ GeV$^2$ in four-gluon
    vertex model. Corresponding upper (yellow) band: $a=1.25$,
    $b=1.67$ GeV$^2$; lower (green) band: $a=1.5$, $b=3.34$
    GeV$^2$. Dashed green line (here also with bands): fully
    iterated results from 3--point complete truncation for the same
    values and ranges of $a$ and $b$.}
\end{figure*}

\begin{figure*}[tb]
  \includegraphics[width=0.49\textwidth]{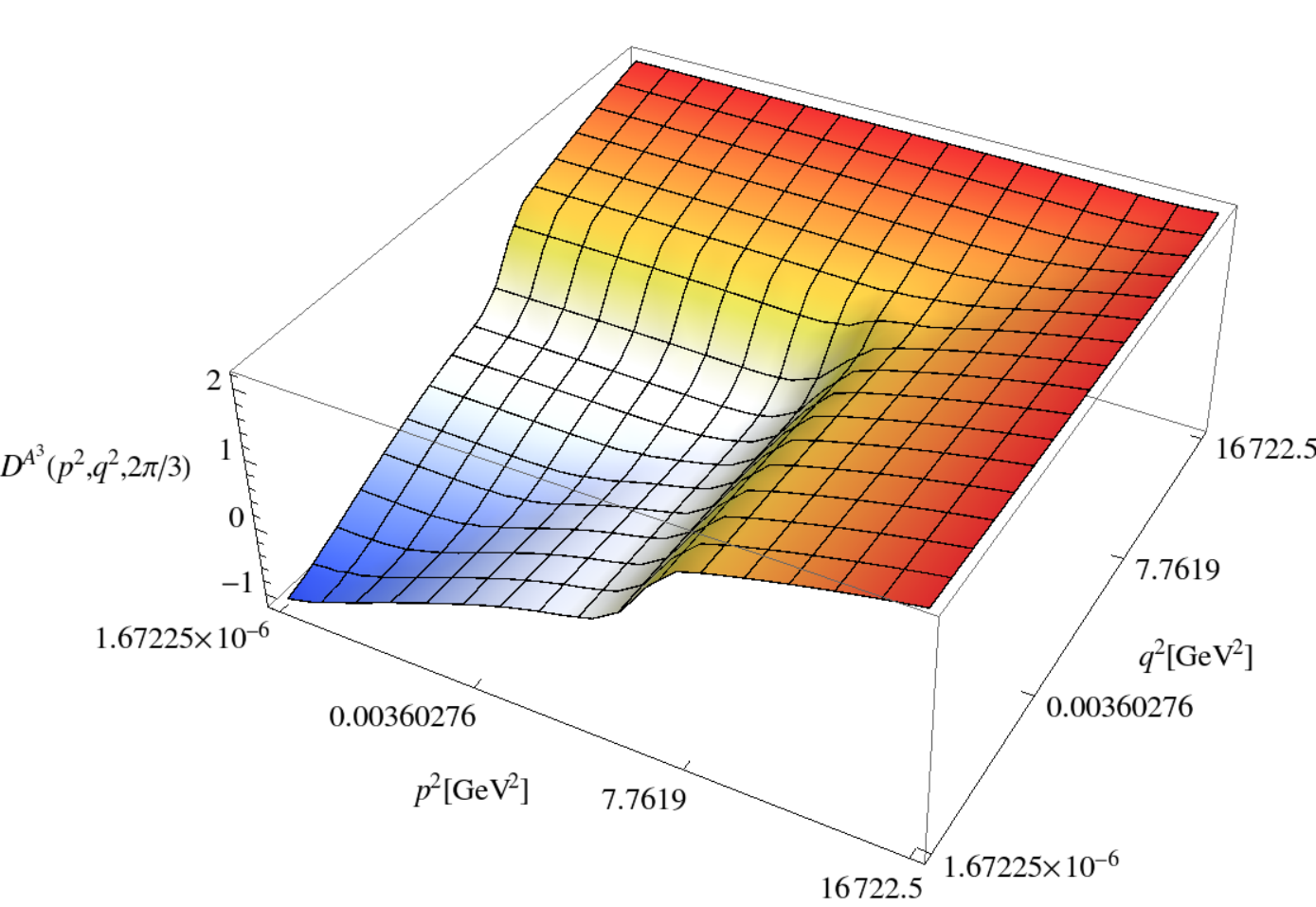}
  \includegraphics[width=0.49\textwidth]{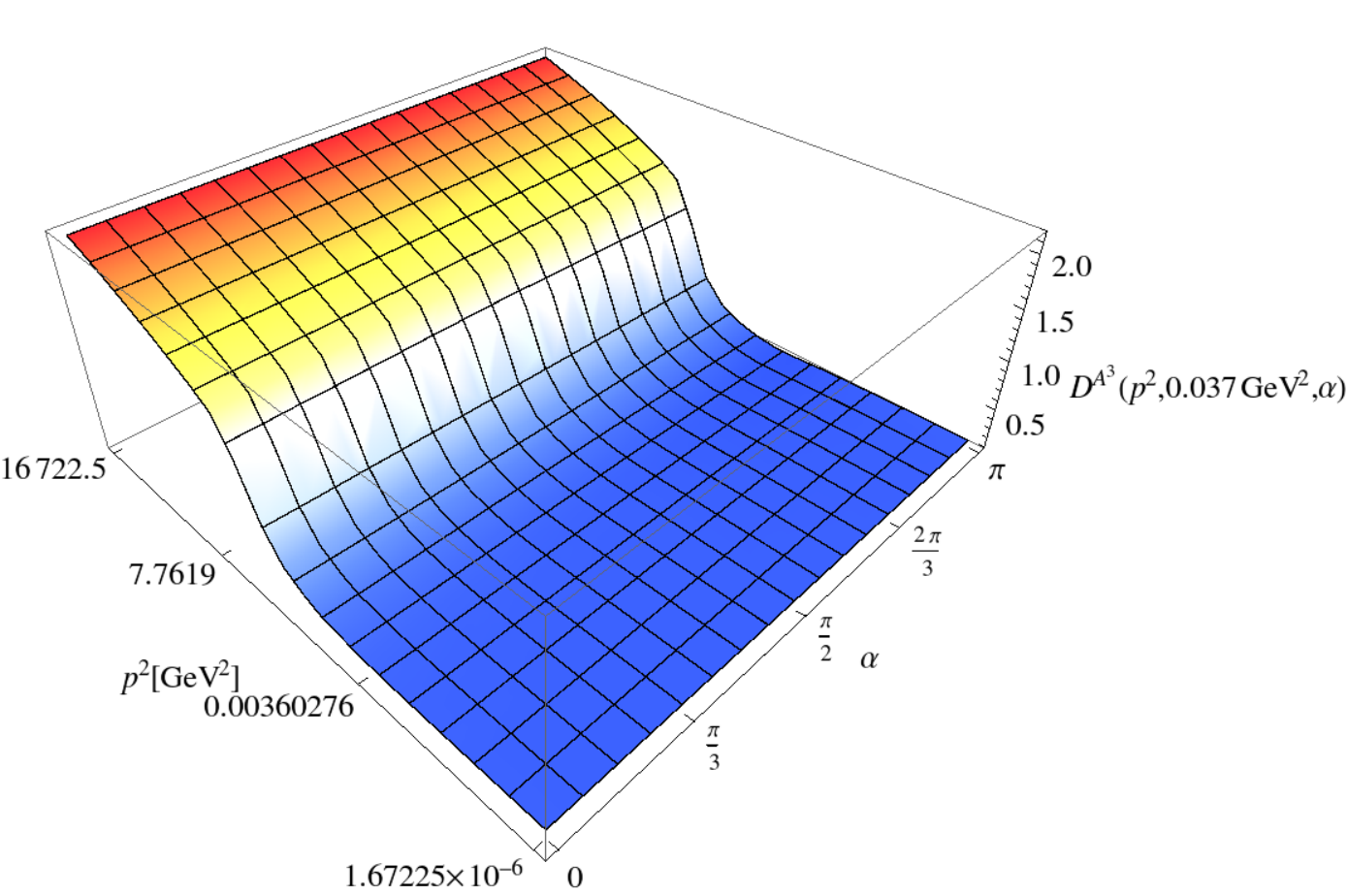}
  \caption{\label{fig:3g_3d-SU2}$SU(2)$ three-gluon vertex
    corresponding to the solid red solution in
    \fref{fig:3g_singleScale-SU2}, but now with two momenta $p$ and
    $q$ at a fixed angle $\alpha = 2\pi/3$ (left), and for fixed
    $q^2 = 0.037\,GeV^2 $ but varying angle $\alpha $
    (right).}
\end{figure*}

Given that lattice data for the three-gluon vertex is only available
for the gauge group $SU(2)$ \cite{Cucchieri:2008qm}, we also
calculated the vertex and the propagators for $SU(2)$. As propagator
input we used a solution of the propagator system with a bare
ghost-gluon vertex and an optimized three-gluon vertex
\cite{Huber:2012kd}, see blue dashed line in
\fref{fig:prop_gl-SU2}. Also in the three-gluon vertex DSE a bare
ghost-gluon vertex was employed, since we know from $SU(3)$, where we
checked this explicitly, that this only leads to minor quantitative
modifications at a level barely visible in the plots presented here.

As for $SU(3)$ we determine three sets of parameters for the
four-gluon vertex model to obtain a band that covers the lattice data,
see lower band in \fref{fig:3g_singleScale-SU2}. These vertex results
were in turn used in the propagator DSEs, where for this calculation a
ghost-gluon vertex calculated in the same manner as the three-gluon
vertex was employed. The resulting propagator dressings are shown in
figs.~\fref{fig:prop_gl-SU2} and \fref{fig:prop_gh-SU2}. The values
for the zero crossings of the three-gluon vertex are $81_{-13}^{+20}$
MeV, $96_{-17}^{+17}$ MeV and $70_{-8}^{+20}$ MeV in this case (configurations in the same order as in \fref{fig:3g_singleScale-SU2}).
The corresponding values for $SU(3)$ are $85_{-7}^{+34}\,MeV$, $100_{-7}^{+15}\,MeV$ and $81_{-10}^{+26}\,MeV$.

For the self-consistent solution we show also a band that corresponds
to the same models of the four-gluon vertex. We note that the band
from the self-consistent calculation is smaller than that from the
three-gluon vertex-only calculations. Furthermore, the bands in the
gluon dressing are smaller for $SU(2)$ than for $SU(3)$. It is
possible that this hints at larger deviations from the real vertex for
$SU(3)$ than anticipated from the close resemblance of $SU(2)$ and
$SU(3)$ propagators. The $SU(2)$ couplings are shown in
\fref{fig:couplings-SU2}. In \fref{fig:3g_3d-SU2} we present
three-dimensional plots of the three-gluon vertex
$D^{A^3}(p^2,q^2,\alpha)$. The shown data corresponds to the red line in
the middle of the bands in \fref{fig:3g_singleScale-SU2}. The plot
with one fixed momentum illustrates that there is only a small angle
dependence in the vertex. The same is true for $SU(3)$. 

\end{document}